%Paper: hep-th/9409128
%From: Josh Boorstein <boo7@bohr.uchicago.edu>
%Date: Wed, 21 Sep 94 10:36:19 CDT

\input harvmac

\def\loop{{\rm loop}}
\def\eff{{\rm eff}}

\def\Tr{{\rm Tr}}

\def\pr{Phys. Rev. }
\def\prl{Phys. Rev. Lett. }

\def\cd{{\cal D}}
\def\ch{{\cal H}}

\def \cl{{\cal L}}

\Title{\vbox{\hbox{EFI-94-42}
\hbox{\tt hep-th/9409128}}}
{{\vbox {\centerline{Wilson Loops, Winding modes and Domain Walls}
\bigskip
\centerline{in Finite Temperature QCD}
}}}

\centerline{\it Joshua Boorstein and David Kutasov}
\smallskip\centerline
{Enrico Fermi Institute}
\centerline {and Department of Physics}
\centerline{University of Chicago}
\centerline{Chicago, IL 60637, USA}
\vskip .2in

\noindent
We discuss the effective action for Polyakov-Wilson loops winding around
compact Euclidean time, which serve as order parameters for the finite
temperature deconfinement transition in $SU(N)$ Yang-Mills gauge theory.
We then apply our results to the study of the high temperature
continuation of the confining phase, and to the analysis of certain
$Z_N$ domain walls that have been argued to play a role in cosmology.
We argue that the free energy of these walls is much larger than
previously thought.

\Date{9/94}
%\draftmode
%

\newsec{Introduction}

There are numerous reasons to study $SU(N)$ gauge theory at finite temperature.
In particular, detailed understanding of the $SU(3)$ theory should be
useful in
describing the physics of QCD in the early universe, and properties of the
quark-gluon plasma $\ref\plasma{J. Cleymans, R.V. Gavai and E. Suhonen, Phys.
Rep. {\bf 130} (1986) 217;
L. McLerran, Rev. Mod. Phys. {\bf 58} (1986) 1021.}$
(the high temperature phase) are important for relativistic
heavy ion collisions.  More formally, one is interested in understanding
the deconfinement transition at large $N$ and study its relation to
the ideas of Hagedorn
$\ref\hag{R. Hagedorn, Nuovo Cimento Suppl. {\bf 3} (1965) 147.}$
and to string theory $\ref\poly{See e.g. A. Polyakov, ``{\it Gauge Fields and
Strings},'' Harwood Academic Publishers (1987).}$.
This is especially interesting since the perturbative
gauge field description, suitable at high temperature and the string one,
valid at low temperature (in the confining phase), should be complimentary
to each other.

As we will briefly review later, the free energy
$e^{-\beta F} = \Tr e^{-\beta H}$ and other physical quantities are
given at finite temperature by a path integral over gauge
fields living on the
Euclidean manifold $R^3 \times  S^1$ with Euclidean time $x^0$ identified with
$x^0+\beta$; gauge fields are periodic: $A_\mu(x^0+\beta,{\bf x})
= A_\mu(x^0,{\bf x})$, while quarks
(if present) are anti-periodic.
There is a nice correspondence between the phase structure of $SU(N)$
gauge fields and $Z_N$ spin systems $\ref\znspin{L. McLerran and B. Svetitsky,
Phys. Rev. {\bf D24} (1981) 450; J. Kuti, J. Polonyi, and K. Szlachanyi,
Phys. Lett. {\bf B96} (1981) 199; B. Svetitsky and L. G. Yaffe,
Nucl. Phys. {\bf B210 [FS6]} (1982) 423.}$.
The high temperature (deconfined) phase
in gauge theory corresponds to the ordered (low temperature) phase of the spin
system and vice versa.
The scalar gauge invariant order parameters which capture the
dynamics of gauge fields are time-like Polyakov -- Wilson loops:
\eqn\wn{W_n({\bf x})= {1\over{N}} \Tr e^{i\int_0^{n\beta}
A_0(x^0,{\bf x}) dx^0}} (with the trace in the fundamental
representation of $SU(N)$).  Only  $N-1$ of the
$W_n$ are independent. Their correlation functions are of utmost
importance; e.g. the quark-antiquark free energy
$F_{q,\bar q}$
defined by:
\eqn\fg{
 e^{-\beta
F_{q,\bar q}({\bf x},{\bf y})}
= \langle W_1({\bf x})W_{-1}({\bf y})\rangle}
measures the free energy of a system with a static quark at
${\bf x}$ and an antiquark at ${\bf y}.$
Higher $W_n$ are related to higher representations of $SU(N)$.
In the confining phase $\langle W_n \rangle =0$ and
$F_{q,\bar q}({\bf x}) \sim |{\bf x}|$ as
$|{\bf x}| \to\infty$, while  in the deconfined phase
 $\langle W_n \rangle \not= 0$ and $F_{q,\bar q}({\bf x}) \sim
{\rm constant}$ as $|{\bf x}| \to \infty$.

To focus on the dynamics of the $W_n$ one may integrate out the other
degrees of
freedom and study an effective action of the general form:
\eqn\seff{\eqalign{S_{\eff} =&
N^2\bigg[\int d^3{\bf x} d^3{\bf y} \sum_nW_n({\bf x}) W_{-n}({\bf y})
G^{(2)}_n({\bf x}-{\bf y})\cr
&+ \int d^3{\bf x_1} d^3{\bf x_2} d^3{\bf x_3}
\sum_{n_1,n_2}W_{n_1}({\bf x_1})W_{n_2}({\bf x_2})W_{-n_1-n_2}({\bf x_3})
G^{(3)}_{n_1,n_2}({\bf x_1},{\bf x_2},{\bf x_3}) + \cdots\bigg].\cr}}
The kernels $G^{(2)},G^{(3)},\ldots$ summarize the dynamics, and are the
object of this paper.  Eq. $\seff$ is the analog of a Landau-Ginzburg
description of the spin system, and can be used
\ref\svet{B. Svetitsky, Phys. Rep. {\bf 132} (1986) 1.}
in the usual way to study
the behavior of $F_{q,\bar q}$ \fg\ and other observables.
At large $N$ the structure
is similar to string theory.  Terms of high order in $W_n$ are suppressed
by powers of $g_{\rm string} = 1/N$.  In particular, it is very interesting to
compare the inverse propagator for Wilson loops
$G^{(2)}$ to the corresponding quantity in conventional
string theories; knowledge of $G^{(2)}$ is sufficient to deduce
the value of the large $N$ deconfinement (Hagedorn) transition ($\beta_H$)
and other
physical properties, like the string tension as a function of temperature.

The immediate motivation for our work is related to three recent ideas:

\noindent
1)  In $\ref\pol{J. Polchinski, \prl {\bf 68} (1992) 1267.}$,
J. Polchinski proposed to study properties of the
confining phase (essentially properties of $G^{(2)}$ \seff) of four
dimensional Yang-Mills theory using perturbation theory, valid
at high temperature, by ``analytically continuing'' the
confining phase to that regime.  These arguments were
since generalized to other cases \ref\kut{D. Kutasov,
Nucl. Phys. {\bf B414} (1994) 33, hep-th/9306013.};
there have also been attempts to match the resulting behavior of $G^{(2)}$
to particular string theories (strings with Dirichlet boundaries
\ref\green{M. B. Green, Phys. Lett. {\bf B282} (1992) 380;
M. Li, Nucl. Phys. {\bf B420} (1994) 339.}, and rigid strings
\ref\pz{J. Polchinski and Z. Yang, Phys. Rev. {\bf D46} (1992) 3667.}).
As pointed out by Polchinski \pol, it
is surprising to obtain properties of an essentially  non-perturbative
object (the confining phase) from
perturbation theory.  Our study of $\seff$ will allow us to clarify
this issue somewhat.

\noindent
2)  It has been proposed $\ref\domw{
K. Kanjantie and L. Karkkainen, Phys. Lett. {\bf B214} (1988) 595;
K. Kanjantie, L. Karkkainen and K. Rummukainen, Nucl. Phys. {\bf B333}
(1990) 100; {\bf B357} (1991) 693;
S. Huang, J. Potvin, C. Rebbi and S. Sanielevici, Phys. Rev {\bf D42} (1990)
2864; [Erratum D43 (1991) 2056];
R. Brower, S. Huang, J. Potvin, C. Rebbi and J. Ross, Phys. Rev. {\bf D46}
 (1992) 4736;
R. Brower, S. Huang, J. Potvin and C. Rebbi, Phys. Rev. {\bf D46}
(1992) 2703.}$,
that at high temperature, when one knows
that the minima of $S_{\eff}$ are at $\langle W_n \rangle =
e^{2\pi i n j\over N}, j=0,1,\ldots,N-1$, the system may possess
domain walls separating regions in space with different
$\langle W_n \rangle$; these domain walls may be of cosmological
interest.  The interface tension (energy per unit area),
$\alpha$,
of such
domain walls
has been argued to be perturbatively calculable and to
go like $\alpha \sim {T^3\over{g(T)}}$
\ref\interf
{T. Bhattacharya, A. Gocksch, C. P. Korthals Altes, and R. D. Pisarski,
Phys. Rev. Lett. {\bf 66} (1991) 998;
Nucl. Phys. {\bf B383} (1992) 497.}\
(with $T$ the temperature and $g$ the running gauge coupling).  If valid,
these arguments could suggest interesting new
phenomena that may have occurred in the early universe
$\ref\cosm{V. Dixit and M. C. Ogilvie, Phys. Lett. {\bf B269} (1991) 353;
J. Ignatius, K. Kajantie, and K. Rummukainen, Phys. Rev. Lett.
{\bf 68} (1992) 737;
C. P. Korthals Altes, K. Lee, and R. Pisarski, hep-ph/9406264.}$.
In addition, these domain walls would correspond to non -- perturbative
effects of order $\exp(-1/g)$ in (finite temperature) gauge theory, which
would
be of more general theoretical interest \ref\shenker{S. Shenker, in
``Random Surfaces, Quantum Gravity and Strings'', Cargese, 1990.}.
However,
various objections to this scenario have been raised
$\ref\diff{V. M. Belyaev, I. I. Kogan, G. W. Semenoff, and N. Weiss,
Phys. Lett. {\bf B277} (1992) 331;
W. Chen, M. I. Dobroliubov, and G. W. Semenoff,
Phys. Rev. {\bf D46} (1992) R1223.}$, $
\ref\dif2{O. A. Borisenko, V. K. Petrov, G. M. Zinovjev,
Phys. Lett. {\bf B264} (1991) 166;
V. V. Skalozub, Mod. Phys. Lett. {\bf A7} (1992) 2895;
O. K. Kalashnikov, Phys. Lett. {\bf B302} (1993) 453;
V. V. Skalozub, Trieste preprint IC/92/405, to appear in Phys. Rev.;
O. A. Borisenko, J. Boh\`a\v cik, V. V. Skalozub, (May, 1994) hep-ph/9405208;
O. K. Kalashnikov, Yukawa Inst. preprint YITP/K-1071 (April, 1994),
hep-ph/9405263.}$.
In particular, it has been claimed $\ref\smilga{
A. V. Smilga, Bern University preprint BUTP-93-03, (May, 1993) and
Santa Barbara preprint NSF-ITP-93-120, (Dec., 1993);
Univ. of Minn. preprint TPI-MINN-94-6-T, (Feb., 1994),
hep-th/9402066.}$ that infrared (IR) divergences
may lead to subtleties in the arguments of $\interf$, although
the leading behavior $\alpha \sim {T^3\over{g(T)}}$ was argued
to be safe.  We will reconsider some of these issues below.

\noindent
3)  In $\kut, \ref\kb{S. Dalley and I. Klebanov, \pr {\bf D47} (1993) 2517;
G. Bhanot, K. Demeterfi and I. Klebanov, \pr {\bf D48} (1993) 4980;
J. Boorstein and D. Kutasov,
Nucl. Phys. {\bf B421} (1994) 263, hep-th/9401044.}$
it has been argued that
two dimensional QCD coupled to adjoint ``quarks'' may be a good
toy model of the relation between large $N$ gauge theory and strings,
exhibiting a non -- trivial spectrum of ``Regge trajectories''
and a large $N$ deconfinement transition.
It is of interest to develop techniques that would allow calculations
of properties of this Hagedorn transition.
The results of this investigation will be reported separately $\ref
\tba{J. Boorstein, to appear.}$.

In this paper we are going to discuss the structure of the effective
action $S_{\rm eff}$, presenting  techniques
to evaluate $G^{(k)}$ perturbatively in the gauge coupling.
The plan of the paper is as follows.  Section 2 is devoted to a brief
review of finite temperature gauge theory, presented
mainly to set the notations and specify the necessary calculations.
In section 3 we describe the calculation of $G^{(2)}$ to one-loop order
and outline the calculations of higher order kernels ($G^{(3)}, \cdots$).
The
perturbative analysis is seen to be infrared finite in a certain
region.  We discuss separately the contributions to $G^{(2)}$ of
gluons and other kinds of adjoint matter.

In sections 4, 5 we discuss the lessons learned from the one-loop
calculations of section 3 for the two problems mentioned above, of
the high temperature limit of the confining phase, and the
interface tension of domain walls.  We show that infrared divergences do not
alter the analysis of $\pol,\kut$ (contrary to recent claims), but
unfortunately one does not appear to be
able to learn much about properties of
the confining phase from this analysis.  On the other hand, for the domain
wall problem infrared issues are found to play a crucial role, and
in fact alter the qualitative behavior of the domain wall
energy per unit area $\alpha\ \interf$, from
$\alpha \sim {1\over g}$ to $\alpha \sim {1\over {g^2}}$.  Section
6 contains a summary of our conclusions and necessary future work.

\newsec{General Formalism}

It is customary \svet,
$\ref\gpy{D. J. Gross, R. D. Pisarski, and L. G. Yaffe, Rev. Mod. Phys.
{\bf 53} (1981) 43.}$ to calculate quantities like
\eqn\zb{Z(\beta) = \Tr e^{-\beta H}}
in the $A_0 = 0$ gauge.  Ignoring matter fields for simplicity, we have
for the QCD Lagrangian $\cl = {1\over{g^2}} \Tr F^2$ the Hamiltonian:
\eqn\ham{\ch = {1\over{2}}\int d^3{\bf x} [g^2 ({\bf E}^a)^2 + {1\over{g^2}}
({\bf B}^a)^2]}
where ${\bf E}^a,{\bf B}^a$ are the color electric and magnetic fields
respectively;  ${\bf E}^a,{\bf A}^a$ are canonically conjugate.  The
physical Hilbert space is spanned by $|{\bf A}(x)\rangle$ satisfying
Gauss' Law constraint
\eqn\gauss{{\bf D}\cdot {\bf E}|{\rm phys}\rangle = 0.}
This constraint may be enforced via a Lagrange multiplier:
\eqn\zp{Z=\int \cd {\bf A}(x)\langle{\bf A}|e^{-\beta \ch} P|{\bf A}\rangle}
where $P = \int\cd \Gamma(x) e^{i\int d^3{\bf x}
\Tr D\Gamma(x)\cdot{\bf E}(x)}$
projects on solutions of Gauss' law \gauss.
Standard Feynman path integral manipulations then lead to the
expression:
\eqn\zpp{Z=\int [\cd A_\mu(x)] e^{-{1\over{g^2}}\int_0^\beta dx^0\int d^3
{\bf x}
\Tr F_{\mu\nu}^2}}
where $\Gamma$ has been renamed $A_0$ and the gauge fields have the periodicity
properties:
\eqn\per{A_\mu(x^0+\beta,{\bf x})=A_\mu(x^0,{\bf x}).}

One can also study more sophisticated questions like what is the free
energy in the presence of sources.  For that one needs to generalize the
Gauss' law constraint \gauss\ to
${\bf D\cdot E} = \rho.$
Repeating the previous discussion one finds \svet, \gpy\ that the free
energy with
static quarks at positions ${\bf x_1},\ldots,{\bf x_n}$ and antiquarks at
${\bf y_1},\ldots,{\bf y_n}$ is given in terms of Euclidean time-like
Wilson loops \wn\ by:
\eqn\fxy{e^{-\beta F({\bf x_1},\ldots,{\bf x_n},{\bf y_1},\ldots,{\bf y_n})} =
\langle\prod_{i=1}^n W_1({\bf x_i})W_{-1}({\bf y_i})\rangle}
with the average performed in the measure \zpp.
We see that the dynamics of the $W_n$ encoded in \seff\ summarizes
many important physical properties
of the theory.

The theory \zpp\ is invariant under gauge transformations
$A_\mu\to U^{-1}A_\mu U + U^{-1}\partial_\mu U$ such that
$U(x^0+\beta,{\bf x}) = z U(x^0,{\bf x}).$  The prefactor $z$ is constrained
by the periodicity of $A_\mu$ \per\ to lie in the center of the gauge group.
For $SU(N)$ the center is simply $Z_N$ and $z=e^{2\pi i n\over N},
n=0,1,\ldots,
N-1.$  Local gauge invariant observables are invariant under these
aperiodic transformations, while $W_n$ \wn\ transforms as:
\eqn\zw{W_n\to z^n W_n.}
The effective action \seff\ must exhibit the global $Z_N$
symmetry \zw, but this symmetry can be broken spontaneously.
Indeed, in the confining (``disordered'') phase where this (``center'') $Z_N$
symmetry is unbroken, $\langle W_n\rangle = 0,$
while at high temperature $\langle W_n\rangle \not= 0$ and $Z_N$ is
spontaneously broken just as in the ordered phase of the spin models \znspin.

To calculate quantities
like \fxy\ in the continuum, one may proceed as follows.
First fix the gauge; \zpp, \per\ does not allow going to $A_0=0$ gauge.  A
convenient gauge choice is
\eqn\gau{\bar A_0^{ab}(x^0,{\bf x})
 = {2\pi\over\beta}\theta_a({\bf x})\delta^{ab}.}
Thus $\bar A_0$ is diagonal and independent of $x_0$ and
$\sum_{a=1}^N \theta_a = 0\ (\rm mod\; 1).$
Then integrate out the spatial gauge
fields $A_i$ in \zpp, and whatever (adjoint) matter fields are present, and
find an effective action of the general form \seff\ written in
terms of $\theta_a({\bf x})$ or \wn:
\eqn\wtheta{W_n({\bf x}) = {1\over N}\sum_{a=1}^N e^{2\pi i n
\theta_a({\bf x})}.}

The description in terms of $\theta_a({\bf x})$ is natural
in the deconfined phase (at high temperatures)
where, as we will see, the effective action is minimized for certain
fixed $\theta_a$, so that $\langle W_n\rangle \not= 0.$
Below the deconfinement transition the $\theta_a$ are randomly distributed
and $\langle W_n\rangle = 0,$
thus it is preferable to describe the dynamics in terms of $W_n$.  We will
use perturbation theory in the gauge coupling $g$,
which in principle should be reliable
at high temperatures (up to possible IR divergences) due to the running
of the gauge coupling; nevertheless, we will mostly
write the effective action in terms of $W_n$ \wtheta, in the spirit of
\pol; the Wilson loops seem to be the suitable variables
for discussing the Hagedorn transition \tba, and may teach us something
about high temperature string theory.
To study
the deconfinement transition and/or the confining phase one must use
non-perturbative tools \tba.

\newsec{The One-Loop Effective Action}

We will be mostly interested in two general classes of gauge systems:

\noindent
1)  Two dimensional Yang-Mills coupled to adjoint (bosonic or fermionic)
matter \kut, \kb.  The Lagrangian for scalar matter is
\eqn\ltwo{\cl=(D_\mu \phi)^2 + m^2\phi^2 + {1\over{g^2}}F^2}
where
\eqn\defin{D_\mu\phi = \partial_\mu \phi + i[A_\mu,\phi]
;\;\;\;F=\partial_0 A_1 - \partial_1 A_0 + i[A_0,A_1].}
We choose the gauge \gau; the Faddeev-Popov determinant is $\det(\partial_0
+i\bar A_0).$
Adding to this the effect of integrating out $A_1,\phi$ to one-loop
order we find
\eqn\detmn{e^{-S_{\eff}^{1-\loop}(\bar A_0)} = \left[\det[(\partial_0
+i\bar A_0)^2]\right]^{-{1\over 2}} (\det (-\bar D^2+ m^2))^{-{1\over 2}}
\det(\partial_0 +i\bar A_0).}
with the three determinants on the r.h.s. arising from
integration over $A_1$,
$\phi$ and the ghosts, respectively.
The ghost contribution exactly cancels that of $A_1$ and we are
left with
\eqn\sefff{S_{\eff}^{1-\loop}(\bar A_0)= {1\over 2} \log \det (-\bar{D}^2+m^2)
.}
For constant $\bar A_0$ one can use the results of \gpy,
$\ref\weiss{N. Weiss, Phys. Rev. {\bf D24} (1981) 475;
{\bf D25} (1982) 2667.}$ to
evaluate $S_{\eff}(\bar A_0).$  We will discuss the determinant
for an ${\bf x}$ dependent $\bar A_0$ \gau.

For fermionic (Majorana) adjoint matter one finds
\eqn\sferm{S^{1-\loop}_{\eff}(\bar A_0) = -{1\over 4} \log \det (
-D^2{\bf 1} + F_{\mu\nu}J^{\mu\nu}+m^2{\bf 1})}
where $J^{\mu\nu}={i\over 4}[\gamma^\mu,\gamma^\nu]$ are the Lorentz
generators and $D_\mu = \partial_\mu - iA_\mu.$

\noindent 2)  D-dimensional pure Yang-Mills (with D$>$2).
Thus we start with
\eqn\ld{\cl={1\over {g^2}} F_{\mu\nu}^2.}
Here, it is convenient to use a background field gauge and Feynman
gauge for the quantum fields.  The one-loop effective
action in this gauge is
\eqn\sol{S^{1-\loop}_{\eff}(\bar A_0) = {1\over 2} \log \det
(-D^2{\bf 1} + F_{\mu\nu}J^{\mu\nu}) - \log \det (-D^2)}
with the Lorentz generators
$(J_{\mu\nu})^{\rho\sigma} = i(\eta_\mu^\rho \eta_\nu^\sigma -
\eta_\nu^\rho\eta_\mu^\sigma).$  The second term on the right hand side of
\sol\ is due to the ghosts, whose contribution is minus that of a
(complex) massless scalar in the adjoint representation.

The one-loop effective actions \sefff, \sferm, \sol\ should be
evaluated for $A_\mu =\bar A_0 \delta_{\mu 0}$ \gau\ and
added to the classical action
$S_{\eff}^{\rm clas} = {\beta\over{g^2}}\int d^{D-1}{\bf x}
(\nabla \bar A_0)^2.$

Most of the analysis will be done for the case of scalar adjoint
matter, which (at one-loop) has most of the qualitative features
of the other two cases.  The properties special to the other cases will be
mentioned later \foot{The following two subsections are rather
technical. Readers interested only in the results may proceed directly to
eq. (3.34).}.

\subsec{The one-loop effective action for scalar adjoint matter -
general considerations }

It is useful to use a first quantized representation to calculate
the determinant \sefff\ (see e.g \ref\strass{M. Strassler, Nucl. Phys.
{\bf B385} (1992)
145.}\ for a recent review).  The
effective action is given by a path integral over worldline trajectories
$x_\mu(t), 0\leq t\leq T$ that wind $n$ times around the compact $x_0$
direction:
\eqn\bound{{\bf x}(t+T) = {\bf x}(t)
;\;\;\;x_0(t+T) = x_0(t) +  n \beta.}
One has also to sum over the length of the world line $T\in{\bf R}^+$
and winding number $n\in {\bf Z}$.
Thus the one-loop effective action
\sefff\
can be written as:
\eqn\sl{S^{1-\loop}_{\eff}[\bar A_0] =
-{1\over 2}\sum_{n=-\infty}^\infty \Tr\int^\infty_0
{dT \over {T}}{\cal N}\int [{\cal D}
x(t)] \exp\left[-\int_0^T dt({1\over{4}}\dot x^2 - iA_0[x(t)]\dot x_0)-m^2 T
\right]}
with the trace taken in the adjoint representation and $\cal N$ a
normalization factor.  This form of
expressing the determinant \sefff\ is related to the standard
Feynman diagram representation by Poisson resummation.
It is convenient to exhibit the $n$ dependence
explicitly by taking $x_0(t)\to x_0(t) + {n \beta t\over T}$
such that $x_0(t)$ is now periodic and the effective action
becomes:
\eqn\scor{\eqalign{S^{1-\loop}_{\eff}[\bar A_0] =&-{1\over 2}
\sum_{n=-\infty}^\infty \Tr\int^\infty_0
{dT \over {T}}{\cal N}\int [{\cal D}x(t)]
\cr &\exp\left[-\int_0^T dt\left({1\over{4}}\dot x^2 +
{n^2 \beta^2\over {4T^2}} - iA_0[x(t)]\left(\dot x_0(t) +{n \beta \over {T}}
\right) \right)-m^2T\right].\cr}}

The path integral over $x(t)$ for a general $A_0$ \gau\ defines a complicated
quantum mechanics problem (which is
closely related to the amplitude for pair production
in a general external electric field),
but fortunately we do not need the full solution,
at least when trying to set up the expansion \seff\
in powers of $W_n$.  As a warm up exercise, consider
the simple case of constant $A_0$ (or $\theta_a$ \gau).
Due to the periodicity of $x_0$ we have:
$$\int_0^T dt A_0\left(\dot x_0(t) +{n \beta \over {T}}
\right) = n \beta A_0,$$ and (replacing $SU(N)$ by $U(N)$,
which is unimportant at
large N):
\eqn\seconst{\eqalign{\cl^{1-\loop}_{\eff}(\theta_a) =& -{1\over 2}
{1\over{{(4\pi)}^{d\over 2}}}\sum_{a,b=1}^N\sum_{n=-\infty}^\infty
\int^\infty_0 {dT \over {T^{1+{d\over{2}}}}}
e^{-{n^2 \beta^2\over {4T^2}} -m^2 T}e^{2\pi i n(\theta_a-\theta_b)}\cr
=&
-N^2\sum_n c_n W_n W_{-n}}}
where $c_n= \left({m\over{2\pi n\beta}}\right)^{d\over 2}
K_{d\over 2}(n m\beta)$
and $K_n$ is a modified Bessel function.
For $m=0, d=4$ this gives the result of \gpy\ (for constant $A_0$ there
is no difference between the scalar \sefff\ and the gauge \sol\ determinants),
while for $d=2$ it agrees with \kut.  We also notice that the winding
of the adjoint scalar around compact time
$n$ in \scor\ is identical to the winding of the Wilson line
$n$ in \seff\ (using \wtheta).
This is a general feature of the one-loop determinants (as we will see below)
which will make
things easier later.  Note also that for ${\it constant}$ $A_0$ (or $W_n \wn$)
the effective action \seff\ is exactly quadratic to one-loop
order, i.e. $G^{(3)}=G^{(4)}=\ldots=0.$  We will see soon that this is not the
case for non-zero momentum and for higher orders in the gauge coupling.

For arbitrary $A_0(x)$ one may proceed as follows.  Write
\eqn\decomp{x(t)=x_{cl}+x_q(t)}
where $x_{cl}$ is constant, and $x_q$ is
the fluctuating quantum mechanical variable in
\scor, which is Gaussian with:
\eqn\correl{\langle x_q^\mu(t_1) x_q^\nu(t_2)\rangle =
 - G(t_1-t_2)g^{\mu\nu}
;\;\;\;G(t+T)=G(t) =|t| - t^2/T.}
Now substituting \decomp\ in \scor\ and expanding (for simplicity
we take $A_0$ to depend on only one of the $d-1$ spatial directions;
no generality is lost because of rotational invariance) we get:
\eqn\expan{A_0(x_{cl}+x_q)=A_0(x_{cl})+A_0'(x_{cl})x_q+
{1\over 2}A_0''(x_{cl})x_q^2+\cdots+{1\over{n!}}A_0^{(n)}(x_{cl})x_q^n
+\cdots.}
We will write this compactly as $A_0[x(t)]=A_0[x]+\tilde{A_0}[x(t)]$ where we
use $x$ to denote $x_{cl}$ and $x(t)$ to denote $x_{cl}+x_q$.

Now expand \scor\ in powers of $\tilde{A_0}$ (or $\theta_a$)
and use \correl\ and \expan\ to average over
quantum fluctuations of the trajectory.
We find that $S^{1-\loop}_{\eff}[A_0]$ is naturally expressed in terms of
bilinears in quantities of the form ($\theta_a^{(n)}$ denotes the $n$'th
derivative of $\theta_a$):
\eqn\compl{V^{(n)}_{n_1,\ldots,n_k} = {1\over N} \sum_{a=1}^N
\theta_a^{(n_1)}\cdots\theta_a^{(n_k)}e^{2\pi i n \theta_a(x_{cl})};\qquad
n_i\geq 1,}
i.e.
\eqn\scomp{\cl^{1-\loop}_{\eff}[V] = N^2 \sum_n
a_{\{n_i\}\{m_j\}}V^{(n)}_{\{n_i\}}\bar{V}^{(n)}_{\{m_j\}}.}

This structure
follows immediately from properties of the trace in
the adjoint representation (for $U(N)$).
A Wilson loop in the adjoint representation of $U(N)$ can
be written as a product of a Wilson loop in the fundamental $N$
 and one in the
anti-fundamental $\bar N$ representation, i.e.
\eqn\snbn{\eqalign{\cl^{1-\loop}_{\eff} =& -{1\over 2}
{1\over{{(4\pi)}^{d\over 2}}}\sum \int_0^\infty
{dT\over{T^{{d\over 2}+1}}} e^{-{n^2\beta^2\over {4T}}-m^2 T}
\bigg\langle \Tr_{\rm adj}
\exp\left[i \int_0^T dt A_0\cdot\left(\dot x_0 + {n\beta \over T}\right)\right]
\bigg\rangle\cr
=& -{1\over 2}{1\over{{(4\pi)}^{d\over 2}}}\sum \int_o^\infty
{dT\over{T^{{d\over 2}+1}}} e^{-{n^2\beta^2\over {4T}}-m^2 T}\cr
&\bigg\langle \Tr_{\rm N}
\exp\left[i \int_0^T dt A_0\cdot\left(\dot x_0 + {n\beta \over T}\right)\right]
\Tr_{\rm N}
\exp\left[i \int_0^T dt A_0\cdot\left(\dot x_0 + {n\beta \over T}\right)
\right]^\dagger\bigg\rangle.\cr}}
Now for the gauge choice \gau, each Wilson loop in
the fundamental representation
can be written in the form \compl\ by first expanding the gauge field as
in \expan\ and then expanding the exponent in powers of $\tilde\theta$, i.e.
\eqn\comex{\eqalign{&\Tr_N \exp \left[i\int_0^T dt A_0
\cdot\left(\dot x_0(t) + {n\beta \over T}\right)\right]
=\sum_{a=1}^N e^{2\pi i n \theta_a(x)}e^{{2\pi i\over\beta}
\int_0^T dt \tilde\theta_a(x(t))\left(\dot x_0(t) + {n\beta \over T}\right)
}\cr
=&\sum_{a=1}^N e^{2\pi i n \theta_a(x)}
\bigg[1 + {2\pi i\over\beta}
\int_0^T dt \tilde\theta_a(x(t))\left(\dot x_0(t) + {n\beta \over T}\right)\cr
&-{4\pi^2\over{\beta^2}}\int_0^Tdt_1\int_0^{t_1}dt_2
\tilde\theta_a(x(t_1))\tilde\theta_a(x(t_2))
\left(\dot x_0(t_1) + {n\beta \over T}\right)
\left(\dot x_0(t_2) + {n\beta \over T}\right)+\cdots\bigg]\cr
=&\sum_{a=1}^N e^{2\pi i n \theta_a(x)}
\bigg[1 + {2\pi i\over\beta }
\int_0^T dt\left(
\theta_a'(x)x_{q}(t)\left(\dot x_0(t)+{n\beta\over T} \right)
+ \cdots\right)\cr
&-{4\pi^2\over{\beta^2}}\int_0^Tdt_1\int_0^{t_1}dt_2\left(
\theta_a'(x)\theta_a'(x)x_q(t_1)x_q(t_2)
\left(\dot x_0(t_1) + {n\beta \over T}\right)
\left(\dot x_0(t_2) + {n\beta \over T}\right)+\cdots\right)\cr
&+\cdots\bigg]\cr
=&N\bigg[W^{(n)} + {2\pi i\over\beta }V^{(n)}_1\int_0^T dt\left(
x_{q}(t)\left(\dot x_0(t)+{n\beta\over T} \right)\right)\cr
&-{4\pi^2\over{\beta^2}}V^{(n)}_{1,1}\int_0^Tdt_1\int_0^{t_1}dt_2\left(
x_q(t_1)x_q(t_2)
\left(\dot x_0(t_1) + {n\beta \over T}\right)
\left(\dot x_0(t_2) + {n\beta \over T}\right)\right)+\cdots\bigg].\cr}}
This expansion can be thought of as an expansion
around slowly varying $A_0$.
One obtains a similar term for the anti-fundamental
Wilson loop and together they give a one-loop effective action which is
written in terms of bilinears of $V_{\{n_i\}}$ \scomp.

It is not clear at first sight how to rewrite the complicated functions
$V_{\{n_i\}}$ \compl\ in terms of the Wilson loops $W_n(x)$ \wtheta,
which is a necessary step for constructing
the expansion \seff, although it is clear
that up to global issues the $W_n$'s exhaust the degrees of freedom
of the $\theta_a$ out of which the $V_{\{n_i\}}$ are constructed.
It turns out that one can
construct an expansion of $V_{\{n_i\}}$ in a power
series in $W_n$.  To leading order in $W_n$ the results are simple;
in fact one can show that
for $k\geq 2$ in \compl,
\eqn\expa{V^{(n)}_{n_1,\cdots,n_k}=O(W^k);}
whereas for k=1 we have:
\eqn\expb{V_{n_1}^{(n)} = {1\over {2\pi i n}} \left({\partial\over{\partial
x}}\right)^{n_1} W_n(x) + O(W^2).}

The terms of order $W^l$ $l\geq2$ above are of course calculable as well.
We will not derive \expa, \expb\ here. To
illustrate the flavor of the arguments, we consider
the simplest non-trivial case of $V_{\{n_i\}}$ with $\sum_{i=1}^k n_i = 2.$
There are in this case only two functions
\eqn\voneone{\eqalign{\tilde V_2^{(n)}\equiv &Z_n={1\over N} \sum_{a=1}^N
2\pi i n \theta_a''e^{2\pi i n \theta_a}\cr
V_{1,1}^{(n)}\equiv &L_n={1\over N} \sum_{a=1}^N
(\theta_a')^2 e^{2\pi i n \theta_a}.\cr}}
Clearly \wtheta
\eqn\deriv{W_n'' = Z_n - (2\pi n)^2 L_n.}
To illustrate \expb\ we have to show that $L_n = O(W^2).$

To establish that, consider
\eqn\xn{\eqalign{0=X_n& = {1\over{N^2}}\sum_{a,b=1}^N\sum_{k=-\infty}^\infty
(\theta_a'-\theta_b')^2 e^{2\pi i n \theta_a-2\pi i k(\theta_a-\theta_b)}\cr
&=2(L_n+L_0 W_n) + 2\sum_{k\not=0,n}[L_kW_{n-k} +
{1\over{4\pi^2k(n-k)}}W_k'W_{n-k}'].\cr}}
All the $L_n$ are expected to be of the same order in $W$; we see from
\xn\ that this order must be $O(W^2)$ and that:
\eqn\ln{L_n = -\sum_{k\not=0,n}{1\over{4\pi^2k(n-k)}}W_k'W_{n-k}'+O(W^3).}
Similar considerations allow one to prove \expa, \expb\ for all ${n_i}$.

The importance of the observations \expa,
\expb\ is that to select terms in \scomp\ which
are bilinear in $W$ (and thus contribute to $G^{(2)}$ \seff),
we need to keep only terms with
$k=1$ \expb\ \foot{Except for the $n=0$ contribution which will be discussed
separately later.}.
In terms of the expansion
in $\tilde \theta$ (see \comex)
this implies that one need only take terms up to
first order in $\tilde \theta$ from each fundamental trace (or up to second
order in $\tilde \theta$ if the trace is done in the adjoint representation).
Higher order (in $W_n$) contributions to the effective action \seff\
can be derived systematically by using \xn, \ln\ and generalizations to
expand the $V_{\{n_i\}}$ in a power series in $W_n$.
The fact that the winding $n$ contribution to the one-loop effective action
\scor\ is directly related to $G_n^{(2)}$ in \seff\ together with
the above observations allow one to calculate $G_n^{(2)}$ to this order.
We now turn to this calculation.

\subsec{Calculation of the inverse propagator for Wilson loops $G_n^{(2)}$}

Starting from \snbn, we have to expand
in $\tilde \theta$, average over the fluctuating $x_q$ and
rewrite the results in terms of bilinears in $W_n$.  Using the results of
the previous subsection it is easy to see that:
\eqn\ac{\eqalign{S^{1-\loop}_{\eff}[\theta]
 &= -{1\over 2}{1\over{{(4\pi)}^{d\over 2}}}
\sum_{n=-\infty}^\infty \sum_{a,b=1}^N\int^\infty_0
{dT \over {T^{1+{d\over{2}}}}}e^{-{n^2 \beta^2\over {4T}} - m^2 T + 2\pi i n
\theta_{ab}(x)}\cr
&\bigg\langle 1+{8\pi^2\over{\beta^2}}
\int_0^T dt_1\int_0^{t_1} dt_2\left(\tilde{\theta_a}[x(t_1)] \tilde{\theta_b}
[x(t_2)]
\left(\dot x_0(t_1) +{n \beta \over {T}}\right)\left(\dot x_0(t_2) +
{n \beta \over {T}}\right) \right)\bigg\rangle\cr}}
where we defined $\theta_{ab}\equiv\theta_a-\theta_b$ and neglected $O(W^3)$
terms.

Now we need to expand $\tilde{\theta}[x(t_i)]$ \expan\ and average
over the $x_q$.
Only those terms with equal
numbers of derivatives in the expansion of the $\tilde\theta_a\tilde\theta_b$
contribute after averaging over quantum fluctuations:
\eqn\ad{\eqalign{&S^{1-\loop}_{\eff}[\theta] =-{1\over 2}
{1\over{{(4\pi)}^{d\over 2}}}\sum_{n=-\infty}^\infty
\sum_{a,b=1}^N
\int^\infty_0 {dT \over
{T^{1+{d\over{2}}}}}e^{-{ n^2 \beta^2\over {4T}} -m^2 T +
2\pi i n\theta_{ab}(x)}\cr
&\left(1+{8\pi^2\over{\beta^2}}
\int_0^T dt_1 \int_0^{t_1} dt_2 \left(\sum_{l=1}^\infty
{(-1)^l\over{l!}}
\theta_a^{(l)}(x)\theta_b^{(l)}(x)
G(t_1-t_2)^l(\ddot G(t_1-t_2) + {n^2\beta^2\over{T^2}}
)\right)\right)\cr}}
where both derivative in $\ddot G$ are with respects to $t_1$.
Note that the $n=0$ contribution to the sum vanishes, since
$\sum \theta_a(x) = 0\; ({\rm mod}\; 1).$

We can shift the variables of integration to make the integral
independent of $t_2$ due to the periodicity of the propagators
$G$.
Doing the
remaining integral over $t_1$ gives a factor of $T^l{l! l! \over {(2l+1)!}}$:
\eqn\ae{\eqalign{S^{1-\loop}_{\eff}[\theta] =
&-{1\over 2}{1\over{{(4\pi)}^{d\over 2}}}\sum_{a,b=1}^N\sum_{n\not=0}
\int^\infty_0 {dT \over {T^{1+{d\over{2}}}}}e^{-{n^2 \beta^2\over {4T}}
-m^2 T + 2\pi i n\theta_{ab}(x)}\cr
&\left(1+{8\pi^2T^2\over{\beta^2}}\sum_{l=1}^\infty {(-T)^l l!\over{(2l+1)!}}
\theta_a^{(l)}(x)\theta_b^{(l)}(x)
(\left({ n\beta\over{T}}\right)^2 - {2\over{T}})\right).\cr}}

Next we change variables to $u={2m^2\over\alpha}T$
where $\alpha=|n| \beta m.$
The last equation can easily be written in terms of the $W_n$
by using \expb.
For convenience we write the answer in momentum space (an integral
over momenta is implied):
\eqn\af{\eqalign{S^{1-\loop}_{\eff}[W] &= -{1\over 2}
{1\over{{(4\pi)}^{d\over 2}}}\sum_{n\neq 0}
\left({2m^2\over{\alpha}}\right)^{d\over{2}}
\int^\infty_0 {du \over {u^{1+{d\over{2}}}}}
e^{-{\alpha\over{2}}(u+{1\over{u}}) }W_n(k)W_{-n}(-k)\cr
&\left(1-2\sum_{l=1}^\infty {(-1)^l l!\over{(2l+1)!}}
\left({k^2\alpha u\over{2 m^2}}\right)^l
({u\over\alpha} - 1)\right).\cr}}

The integral over $u$
can be expressed in terms of modified Bessel functions using
the defining equation
$$K_\nu(x)=
{1\over{2}}\int_0^\infty du u^{\nu-1} e^{-{x\over{2}}(u+{1\over{u}})
}.$$
{}From this we obtain
\eqn\ag{\eqalign{S^{1-\loop}&_{\eff}[W] = -
\sum_{n\neq 0}
\left({m^2\over{2\pi\alpha}}\right)^{d\over{2}}W_n(k)W_{-n}(-k)\cr
&\left(K_{d\over{2}}(\alpha) +
2\sum_{l=1}^\infty {(-1)^l l!\over{(2l+1)!}}
\left({\alpha k^2\over{2 m^2}}\right)^l
(K_{l-{d\over{2}}}(\alpha) -
 {1\over \alpha}K_{l-{d\over{2}}+1}(\alpha))\right).\cr}}

As mentioned above,
to complete the calculation we have to evaluate the zero winding, $n=0$,
contribution.  For constant $\theta_a$ the zero winding sector
gives rise to an
unimportant temperature independent constant, which is usually dropped.
In general, however, the $n=0$ contribution is non -- trivial.
Indeed, to quadratic order in $W_n$ one finds:
\eqn\ai{S^{1-\loop}_{\eff}[\theta]_{n=0} =
-{1\over 2}{1\over{{(4\pi)}^{d\over 2}}}\sum_{a,b=1}^N
\int^\infty_0 {dT \over {T^{1+{d\over{2}}}}}e^{-m^2 T }
\left(1+{16\pi^2T\over{\beta^2}}\sum_{l=1}^\infty {(-T)^l l!\over{(2l+1)!}}
\theta_a^{(l)}(x)\theta_a^{(l)}(x)\right)}
where half of the contribution is due to the fundamental and half
to the anti-fundamental representations.

We need to write this in terms of non-zero winding Wilson loops
$W_n$, which
can be achieved by utilizing the derivation of
\ln\ and generalizations.
In momentum space, one finds:
\eqn\aii{S^{1-\loop}_{\eff}[W]_{n=0} =
{1\over{(4\pi)^{d\over 2}}}\sum_{n\not=0}{2m^2\over{\alpha^2}}
W_n(k)W_{-n}(-k)
\int^\infty_0 {dT \over {T^{d\over 2}}}e^{-m^2 T}
\sum_{l=1}^\infty {(-T k^2)^l l!\over{(2l+1)!}} }
where we have dropped a temperature independent constant, and as usual
a momentum integral is implied.

Finally, doing the $T$ integral gives
\eqn\aj{S^{1-\loop}_{\eff}[W]_{n=0} = \left({m^2\over{4\pi}}\right)^{d\over 2}
\sum_{n\neq 0}{2\over{\alpha^2}}
W_n(k)W_{-n}(-k)\left(
\sum_{l=1}^\infty {(-1)^l l! \Gamma(l-{d\over 2}+1)
\over{(2l+1)!}}\left({k^2\over{m^2}}
\right)^l \right).}

We can now combine
\ag\ with \aj\ and find the one-loop contribution
to the quadratic action for Wilson loops:
\eqn\bb{\eqalign{&S^{1-\loop}_{\eff}[W] = -\sum_{n\neq 0}
\left({m^2\over{4\pi}}\right)^{d\over 2}
W_{-n}(-k)W_n(k)
\biggl( \left({2\over\alpha}\right)^{d\over 2} K_{d\over{2}}(\alpha)  \cr
&+\sum_{l=1}^\infty {(-1)^l l!\over{(2l+1)!}}
\left({k^2\over{ m^2}}\right)^l\left[
\left({\alpha\over 2}\right)^{l-{d\over 2}-1}
(\alpha K_{l-{d\over 2}}(\alpha)
-K_{l-{d\over 2}+1}(\alpha)) + {2\over{\alpha^2}}\Gamma(l-{d\over 2}+1)\right]
\biggr).\cr}}
Adding this to the term in the effective action coming from the classical
action we get
$G^{(2)}_n$
to one-loop order:
\eqn\gtol{\eqalign{&G^{(2)}_n(k) = {1\over(4\pi)^{d\over 2}}
{m^2\over{\alpha^2}}{k^2\over{2g^2N}}
-\left({m^2\over{4\pi}}\right)^{d\over 2}
\biggl( \left({2\over\alpha}\right)^{d\over 2} K_{d\over{2}}(\alpha)  \cr
&+\sum_{l=1}^\infty {(-1)^l l!\over{(2l+1)!}}
\left({k^2\over{ m^2}}\right)^l\left[
\left({\alpha\over 2}\right)^{l-{d\over 2}-1}
(\alpha K_{l-{d\over 2}}(\alpha)
-K_{l-{d\over 2}+1}(\alpha)) + {2\over{\alpha^2}}\Gamma(l-{d\over 2}+1)\right]
\biggr)
.\cr}}
Recall that $\alpha=m\beta|n|$. Eqns. \bb, \gtol\ are the main result of this
section.
We now turn to the study of some of their properties.

\subsec{Properties of $G^{(2)}_n(k)$}
Despite appearances, the limit $m\to0$ of \gtol\ is actually regular. To study
this limit it is convenient to
rewrite the sum over Bessel functions in terms of a compact
integral by reversing the order of the sum and integral in \ae.
This gives
\eqn\bc{\eqalign{G^{(2)}_n(k)& = {1\over(4\pi)^{d\over 2}}
{m^2\over{\alpha^2}}{k^2\over{2g^2N}}
-\left({m^2\over{2\pi\alpha^2}}\right)^{d\over 2}
\biggl( \alpha^{d\over 2} K_{d\over{2}}(\alpha)  \cr
&+\int_0^\infty{du\over{u^{1+{d\over 2}}}}
e^{-{1\over 2}(\alpha^2 u +{1\over u})}f({\alpha^2 k^2 u\over{2 m^2}})-
\int_0^\infty{du\over{u^{{d\over 2}}}}
(e^{-{1\over 2}(\alpha^2 u +{1\over u})}-e^{-{1\over 2}\alpha^2 u})
f({\alpha^2 k^2 u\over{2 m^2}})
\biggr)
.\cr}}
where for $x>0$
\eqn\deff{f(x)=-{1\over{\sqrt{x}}}e^{-{x\over 4}}\int_0^x dt{\sqrt{t}\over 2}
e^{t\over 4}}
and for $x<0$
\eqn\defh{f(x)={1\over{\sqrt{|x|}}}e^{{|x|\over 4}}\int_0^{|x|}
dt{\sqrt{t}\over 2}e^{-{t\over 4}}.}

To study the convergence properties of the integrals in \bc\ we have to examine
the asymptotic behavior of $f(x).$  First note that
in both cases $f(x) \sim -x$ as $x\to 0.$
Thus, for $d<4$ the integrals in \bc\ are convergent in the UV
(for small $u$). The second
integral exhibits a standard logarithmic UV divergence which corresponds to
coupling constant renormalization in $d=4$.
For $x>0,$ it is easy to see that
$f(x)$ goes to a constant ($-2$) as $x\to\infty.$
Hence for $k^2>0$ the integral representation for $G^{(2)}$ \bc\
is convergent even for m=0 (and fixed $\beta$), for which the exponential
IR suppression of the integrals (provided by the mass)
is absent. This is to be contrasted
with the infrared divergent results one finds if one expands $f$
in a power series in ${\alpha^2\over m^2}k^2u$
(derivative expansion) and integrates term by term.

In the case $x<0$ as $|x|\to\infty$ the integral in \defh\ converges to
$2\sqrt{\pi}$.  This implies that $f(x)$ grows as
${e^{{|x|\over 4}}\over{\sqrt{|x|}}}$ as $x\to -\infty$.
Putting this asymptotic form for $f(x)$ into \bc\ for $k^2<0$ and
large $u$ gives
\eqn\expg{G^{(2)} \sim \left({m^2\over{\alpha^2}}\right)^{d\over 2}
\sqrt{{m^2\over{\alpha^2 |k|^2}}}
\int^\infty {du\over{u^{{d+3}\over2}}}\exp\left[-{\alpha^2 u\over 2}
\left(1-{|k|^2\over{4m^2}}\right)\right]}
For $k^2<-4m^2$ this integral diverges.  By rescaling the
integration variable we see that
in this range $G^{(2)}$ develops a branch cut:
$G^{(2)}\propto(4m^2-|k|^2)^{(d+1)/2}$.
We will discuss this cut below
and will see that it has some important consequences.

As mentioned above, we find a regular expression for $G^{(2)}$
as $m\to 0$. This is easiest to see from \bc, in the limit $\alpha\to0$,
$\alpha/m=\beta|n|$ fixed. We find:
\eqn\bcc{\eqalign{G^{(2)}_n(k) = &{1\over(4\pi)^{d\over 2}}
{1\over{n^2\beta^2}}{k^2\over{2g^2N}}
-\left({1\over{4\pi n^2\beta^2}}\right)^{d\over 2}
\biggl( 2^{d-1}  \cr
&+\int_0^\infty{du\over{u^{1+{d\over 2}}}}
e^{-{1\over {4u}}}f(n^2\beta^2 k^2 u)-
2\int_0^\infty{du\over{u^{{d\over 2}}}}
(e^{-{1\over {4u}}}-1)
f(n^2\beta^2 k^2 u)
\biggr)
.\cr}}
By using the asymptotic behavior of \deff, \defh\ we see that
$G_n^{(2)}(k^2>0)$
is regular. The branch cut mentioned above starts now at $k^2=0$.

Before going on to apply these results
to physical problems, in the next subsections we briefly comment on similar
calculations
for gauge fields and fermionic matter.

\subsec{Fermionic matter}

For fermionic matter we start from \sferm\ and remembering that
the fermions have anti-periodic boundary conditions, i.e. $p_0=
{\pi\over \beta}(2 n +1),$ we derive the analog of \scor.
Since the $\gamma$ matrices are anti-commuting operators, it is natural in
this first quantized approach to
introduce world line fermions to represent them.  The easiest way to do
this is to implement a supersymmetric generalization of \bound, introducing
superpartners $\psi_\mu(t)$ for the $x_\mu(t)$.
For more detail see \strass.

For adjoint fermionic matter in two dimensions we obtain
\eqn\scorf{\eqalign{&S^{1-\loop}_{\eff}[\bar A_0] ={1\over 4}
\sum_{n=-\infty}^\infty(-1)^n \Tr\int^\infty_0
{dT \over {T}}{\cal N}\int [{\cal D}x(t)][{\cal D}\psi(t)]
\cr &\exp\left[-\int_0^T dt\left({1\over{4}}\dot x^2 + {1\over 2}\psi\cdot
\dot{\psi} + 2i \psi^0 F_{01} \psi^1+
{n^2 \beta^2\over {4T^2}} - iA_0[x(t)]\left(\dot x_0(t) +{n \beta \over {T}}
\right)\right)-m^2T\right]\cr}}
where $F_{01}=\partial_1A_0$.

We need to expand to second order in $A_0$ and then average over
the quantum fluctuations of the path, as well as the $\psi$ fields.
Note that the $\psi$ fields satisfy
\eqn\corrf{\eqalign{\langle\psi^\mu(t_1) \psi^\nu(t_2)\rangle
=&-G_F(t_1-t_2)g^{\mu\nu}\cr
-G_F(t+T,t')=G_F(t,t')=&{\rm sgn}(t-t').\cr}}
After some algebra we obtain
\eqn\adf {\eqalign{S^{1-\loop}_{\eff}[A] &={1\over 4}{1\over{(4\pi)^{d\over2}}}
\sum_{n=-\infty}^\infty (-1)^n \Tr \int^\infty_0 {dT \over
{T^{1+{d\over{2}}}}}e^{-{ n^2 \beta^2\over {4T}} -m^2 T +  i n\beta A(x)}\cr
&\biggl(1-\int_0^T dt_1 \int_0^{t_1} dt_2 \biggl(\sum_{l=1}^\infty {(-1)^l
\over{l!}}
(A^{(l)}(x))^2  [G(t_1-t_2)]^{l-1}(-l G_F(t_1-t_2)^2\cr
&+G(t_1-t_2)(\ddot G(t_1-t_2) + {n^2\beta^2\over{T^2}}))\biggr)\biggr).\cr}}

The rest of the analysis follows straightforwardly from
the bosonic case.  The term introduced by
the world sheet fermions does not effect the position of the cut in
the effective action.

\subsec{Gauge fields}
We write here only the one-loop effective action obtained from
integrating out gauge fields in four dimensions.
The interested reader may consult
the work of Strassler \strass\ for a thorough discussion on
how to handle vector fields in this approach.  The effective action is given by
\eqn\advec{\eqalign{S^{1-\loop}_{\eff}[A] &=-{1\over 4}
{1\over{(4\pi)^{d\over2}}}
\sum_{n=-\infty}^\infty  \Tr \int^\infty_0 {dT \over{T^3}}
e^{-{ n^2 \beta^2\over {4T}}  +  i n\beta A(x)}\cr
&\biggl(1-\int_0^T dt_1 \int_0^{t_1} dt_2 \biggl(\sum_{l=1}^\infty {(-1)^l
\over{l!}}(A^{(l)}(x))^2  [G(t_1-t_2)]^{l-1}\cr
&(-4 l
+G(t_1-t_2)(\ddot G(t_1-t_2) + {n^2\beta^2\over{T^2}}))\biggr)\biggr).\cr}}
The rest of the analysis is similar to the scalar case with
the cut starting at $k^2=0$ as expected.

\newsec{Consequences for the high temperature continuation of the
confining phase}

In reference \pol, J. Polchinski has proposed to use the form
of $G^{(2)}$ \seff\ to study the confining phase of QCD.  The idea, inspired
by string theory, is to view the perturbative calculation of $G^{(2)}$,
$G^{(3)},\ldots$ which is in principle only
valid at high temperature, in the deconfined
phase, as an analytic continuation of the confined
phase to high temperatures.  In string theory such analytic
continuations are routine; the analogs of the Wilson lines $W_n$,
which are winding modes around compact time are governed by an action:
\eqn\str{\cl_{str}= {1\over{g_{st}^2}} W_n({\bf k}) W_{-n}(-{\bf k})
[{{\bf k}}^2 + M_n^2(\beta)]+O(W^3); \qquad M_n^2(\beta) = \beta^2 n^2 - C}
with C a positive constant. There is no difficulty in formally
extending this formula to $\beta < \beta_H = \sqrt{C}.$

It has been pointed out \ref\kogan{ I. I. Kogan,
Princeton University preprint PUPT-1415, (Nov. 1993),
hep-th/9311164.} that calculations such as those of \pol, \kut\
in which $G^{(2)}$ is calculated only at zero momentum may suffer from
IR divergences, since to find $S_{\eff}$ in that case, massless particles
are sometimes integrated out, and the effective action is in principle
non-local
already at ${{\bf k}}^2 \approx 0$.  Of course, in four dimensional
gauge theory the spatial gauge fields that are integrated out are not
really
massless but rather develop a magnetic mass \gpy\ $m\approx g^2(\beta)/\beta$
(which is unfortunately incalculable in perturbation theory), and in adjoint
$QCD_2$
 one may turn on by hand a mass for the matter fields (whose role
is precisely to mimic the above magnetic mass).  One would however like to
understand whether
the results of \pol, \kut\ are reliable in the massless limit, and
more importantly whether one can indeed learn about the confining
phase from such perturbative calculations.

The results of section 3 help to resolve both issues.  Regarding
the IR divergences, we see that they are harmless, even for massless
adjoint matter (gluons) in $2d$ ($4d$).  Consider for example
the case of massless,
bosonic adjoint $QCD_2$.
The inverse propagator $G^{(2)}$ vanishes when \bcc:
\eqn\repl{1 - {k^2\over{4 g^2 N}}+{1\over 2}\int_0^\infty {du\over{u^2}}
e^{-{1\over {4u}}} f(k^2\beta^2 n^2 u)-
\int_0^\infty {du\over{u}}(e^{-{1\over {4u}}}-1)
 f(k^2\beta^2 n^2 u) = 0}
(see section 3 for definition of $f$ and derivation).
In \pol, \kut\ the last two terms on the r.h.s. of \repl\
were neglected, because they are formally small
at high temperature (of higher
order in $\beta^2 g^2$) when the sum of the first two vanishes,
i.e. when $k^2=4g^2N$.
We saw in section 3 that the
IR singularity due to integrating out a massless field results in a
cut to the {\it left} of $k^2 =0$. This cut is not dangerous
at $4g^2N=k^2>0$.  The full inverse propagator \repl\
vanishes at $k^2=4g^2N[1+O(g^2\beta^2)]$ (up to logarithmic
corrections), approaching the result
of \pol, \kut\ as the temperature goes to infinity.
To put it differently, integrating out the massless constituents
does not introduce infrared subtleties since the Wilson loops $W_n$
are tachyonic at high temperature.
Thus the calculations of \pol\ and \kut\
are valid,
contrary to the recent claims \kogan.

Unfortunately, our result \bb\ seems, at
least naively, to invalidate the basic idea of using perturbation theory
for $G^{(l)}$ to study the confining phase of QCD.
Our calculations give a (one loop) glimpse of the analytic structure
of the effective action \seff\ in momentum space.
Indeed, one of the most important
qualitative differences between the confining and deconfined phases of QCD
is the analyticity properties of the Green's functions,
such as $G^{(2)}(k)$.
In the confining, low temperature phase, $G^{(2)}$
is expected to be
an analytic function of $\bf k^2$ to leading order in
$1/N$ (compare to \str), since any singularities would have to
be interpreted in terms of interactions of the $W_n$ among themselves and
with other
singlet bound states and would be down by powers of the string coupling
$1/N$.
In the high temperature phase, one expects the structure to change
drastically.  The $W_n$ are no longer the natural degrees of freedom, and
in particular there are non-singlet operators (quarks, gluons) that may couple
to $W_n$. Thus one expects $G^{(2)}$ to contain branch cuts corresponding to
pair production of such non-singlet degrees of freedom and to have a
complicated
analytic structure typical of the deconfined phase already
in {\it leading order in $1/N$}.

To determine whether the perturbative calculations
of \pol, \kut\ correspond to an analytic continuation from the
confining phase or to properties of the deconfined phase one has
to study the analytic structure of $G^{(l)}$
and in particular look for branch cut singularities signaling
the propagation of constituents.
The cuts we found
in section 3 are precisely of this kind; they seem to correspond to
coupling of $W_n$ to two constituents (quarks, gluons).  Thus, the natural
conclusion from our analysis is that the perturbative calculations
performed in \pol,
\kut\ and section 3 give the effective action for $A_0$ in the $\it deconfined
 \ phase$, written in terms of the variables $W_n$, and not, as one would
hope, the analytic continuation of the confining phase\foot{
In principle one has to investigate the analytic structure of higher
order (in $g$) corrections to \seff, but it is clear physically that
the qualitative picture presented here should persist to all orders.}.
The latter would have a very
different analytic structure than the former, and than what we find (section
3).  One can not easily infer any properties of the confining phase
(other than it being unstable at high temperature) from these calculations.
In particular the number of degrees of freedom of the string does not seem
to be easily extractable.  There does not seem to be a simple
way to calculate in the
confining phase without actually following
$G^{(2)}$ up to the phase transition at $\beta_H$, where all the
non-singlet singularities should disappear.
This requires a non -- perturbative analysis, which may be feasible in
certain toy
models \tba.

\newsec{Consequences for the structure of domain walls}

In the deconfined phase it is natural to express the action in terms
of the $\theta$ variables \gau\ as opposed to Wilson loops.
For slowly varying $\theta_a(x)$, the Lagrangian \seff\ takes the form (to
one loop order):
\eqn\lonel{\cl={4\pi^2\over \beta^2g^2}\sum_{a=1}^N(\nabla\theta_a)^2+
V_{\rm eff}(\theta)}
where the kinetic term comes from the classical Lagrangian \ld\ and the
one loop
potential $V_{\eff}$ is \seconst, which in four dimensions is
(for $0\leq\theta_{ab}\leq 1$):
\eqn\veff{\eqalign{V_{\eff}(\theta)
=& -{1\over{\pi^2\beta^4}}\sum_{a,b=1}^N\sum_{n=-\infty}
^{\infty}{1\over{n^4}}e^{2\pi i n\theta_{ab}}\cr
=&{2\pi^2\over{3\beta^4}}\sum_{a,b=1}^N B_4(\theta_{ab})\cr}}
where $\theta_{ab}=\theta_a-\theta_b$ and $B_4(x)$ is the Bernoulli polynomial
$$B_4(x)=x^4-2x^3+x^2-1/30$$

The Lagrangian \lonel\ is invariant under the $Z_N$ symmetry \zw\
$\theta_a\to\theta_a
+k/N$. This
symmetry implies that there are in fact $N$ different
minima of the potential $V_{\eff}$, corresponding to all
$\theta_a = k/N$ for $k=0,1,\ldots,N-1.$  In terms of Wilson loops,
these minima correspond to $W_n = e^{2\pi i n k\over N}$.

By analogy with spin systems \znspin, it is natural to study domain
walls separating regions in space corresponding to
different vacua of $V_{\eff}$\foot{There is some debate in the literature
regarding the existence and physical significance of such walls
\domw -- \cosm,
\diff -- \smilga. We will assume that they exist and study
some of their features.}.
In particular, one can attempt to calculate the interface energy
$\alpha$, defined
as the free energy per unit area of the wall.

At high temperature, when $g(\beta)$ is small, one may hope to use
semiclassical
techniques to study these walls \interf. To calculate the free energy
of a domain wall
between  regions in space corresponding to
$W_n=1$ and $W_n=\exp(2\pi in/N)$, say, we have to find
a solution
of the effective action \lonel\ which behaves as $\theta_a(z\to-\infty)=0$,
$\theta_a(z\to\infty)=1/N$ (for all $a$).
It can be shown \interf\ that the minimal action solution is obtained
by choosing a particular path in $\theta$ space, parametrized
by\foot{Note that
in this section $N$ is not assumed to be large.}:
\eqn\para{\eqalign{\theta_a=&q(z)/N,\qquad a=1,\ldots,N-1,\cr
\theta_N=&-{N-1\over N} q(z).\cr}}
In this parameterization the Wilson loop takes the form
$$W_n = e^{2\pi i n q\over N}(1+{1\over N}(e^{2\pi i n q}-1)).$$
Thus $q=0$ corresponds to $W_n = 1$ while $q=1$ corresponds
to $W_n = e^{2\pi i n \over N}$.
The action \lonel\ for $q(z)$ \para\ is ($L_t^2$ is the
transverse area of the domain
wall):
\eqn\sbeff{S_{\rm eff}={L_t^2 4(N-1)\pi^2\over\beta}
\int dz\left( {1\over{g^2 N}}(q')^2 + {1\over 3\beta^2}[q]^2(1-[q])^2\right).}
$[q]\equiv q\;{\rm mod} \;1$.
The solution with the right boundary conditions is
\eqn\solut{q(z)={\exp(\sqrt{N\over 3}gz/\beta)\over1+
\exp(\sqrt{N\over 3}gz/\beta)}}
Plugging it back into \sbeff\ we find that the interface tension is
\eqn\albub{{S_{\eff}\over\beta L_t^2}=\alpha =
{4(N-1)\pi^2\over{3\sqrt{3N}}}{1\over \beta^3g}.}

It is clear from \solut\ that the scale of variation of the solution $q(z)$ is
$\sim 1/g$. Hence, each derivative comes with a power of $g$.
Formally, this implies
that higher derivative terms in the effective action (as well as higher loop
contributions) modify the solution \solut\ and $\alpha$ \albub\ only slightly,
as the effective coupling at high temperature $g(\beta)$ is small.
One might worry
\smilga\ that infrared effects due to integrating out
massless gluons may spoil the formal power counting.
To examine this issue one has to look at higher derivative terms in the
effective
action for $q(z)$.

Our results are in general insufficient for this task, since
we have not calculated $S_{\eff}$ for arbitrary $q(z).$  However, to study
the above infrared issues it is enough to consider the behavior of the
domain wall profile $q(z)$ as $z\to-\infty$ (say), since then
$q$ \solut\ is small, $q(z)\simeq\exp(\sqrt{N\over3}{gz\over\beta})$,
and one can use our results from
section 3 to write the effective action to order $q^2$ (but to all
orders in the derivative expansion \expan)\foot{The modifications
to eq. (5.7) as compared to eq. \bcc\ are
due to the difference between the scalar \sefff\ and gauge
\sol\ determinants
(see section 3), and should not matter for the qualitative remarks that
follow.}:
\eqn\sefth{\eqalign{S_{\eff}(q) =& {L_t^2 4(N-1)\pi^2\over\beta}
\int dz q(z)\bigg(- {\partial_z^2\over{g^2N}} + {1\over 3\beta^2}\cr
+&{1\over{2\pi^2\beta^2}}
\sum_{n\not=0}{1\over{n^2}}\bigg({1\over 4}\int_0^\infty {du\over{u^3}}
e^{-{1\over {4u}}} f(-\partial_z^2\beta^2 n^2 u)-
{1\over 2}\int_0^\infty {du\over{u^2}}(e^{-{1\over {4u}}}-1)
 f(-\partial_z^2\beta^2 n^2u)\cr
-&n^2\beta^2\partial_z^2\int_0^\infty {du\over u}(e^{-{1\over {4u}}}-1)
 (f(-\partial_z^2\beta^2 n^2u)+1)\bigg)\bigg)q(z)+O(q^3,\cdots).\cr}}
We see that the situation is different from that of section 4. Plugging in
the asymptotic form of \solut\ in \sefth\ we find that $k^2=-\partial_z^2=
-(N/3)g^2/\beta^2<0$, so the terms on the second and third lines of \sefth\
are not small corrections to the ``leading behavior'' obtained from the
first two terms
on the r.h.s. Unlike the $W_n$ in section 4, the mass squared
of $q$ is positive (and equal to the square of the electric mass).
Therefore, the branch cut discussed in section 3 significantly alters the
behavior coming from \lonel. It seems at first sight that \sefth\ completely
eliminates the possibility of the existence of a domain wall, since
$q(z)=\exp(az)$ is not a solution for any real $a$. However,
the correct interpretation is different.

One important
effect due to higher order contributions in $g$ that we have neglected
so far is the generation of a ``magnetic mass'' for the spatial components
of the gauge field (static magnetic fields are screened). This magnetic
mass, which is of order $m_{\rm mag}\simeq g^2/\beta$ is not perturbatively
calculable
\ref\linde{A. Linde, Phys. Lett. {\bf 96B} (1980) 289.},
\gpy\
but its presence alters the picture following from \sefth. As we saw in section
3, it shifts the branch cut from
$k^2=0$ to $k^2=-4m_{\rm mag}^2\simeq-4g^4/\beta^2$.
The scale of the solution \solut\ of \interf\ is on the other hand
the ``electric mass''
$m_{\rm el}\simeq g/\beta$.
Since $g$ is small in high temperature QCD, $m_{\rm mag}
\ll m_{\rm el}$, so that even with the modification of \sefth\ due to the
magnetic mass, the solution \solut\ is not infrared stable. However,
as is clear from \bc\ (with $m=m_{\rm mag}$, and an appropriate
generalization for the gauge case), there is now a solution $\tilde q(z)$
which behaves asymptotically as $\tilde q(z)\simeq \exp(az)$ with
$a\simeq m_{\rm mag}$.

The main question now is whether the scale of variation
of the domain wall solution remains $m_{\rm mag}$ for moderate $q$ as well.
In \smilga\ it has been argued that the answer to this question is
negative since the infrared scale is determined by the mass of
the space -- like gluons ${\bf A}_{ab}$ which for generic $\theta_a$
is $\sqrt{m_{\rm mag}^2+({\theta_{ab}\over\beta})^2}$.
In that case, the form of the domain
wall solution $q(z)$ \solut\ is only significantly modified at the tails
$|z|\to\infty$, and
the leading behavior of the interface tension \albub\ is not effected \smilga.
However,
we believe that the scale of variation of the
solution $q(z)$ is of order $m_{\rm mag}$ throughout the wall.
The point is that there are many physical
states whose masses are of order $m_{\rm mag}$ for generic
$\theta$ (or $q$). Examples include the space -- like gluons
${\bf A}_{ab}$ with $a=b$, and gauge invariant combinations
like $Tr{\bf F}^2$. These can be pair produced at higher
orders in the loop expansion and lead, as explained above,
to a domain wall profile which varies on the scale $m\approx m_{\rm mag}$
for all $q$.

Thus, we conclude that infrared effects
change the scale of the domain wall solution of \interf\ from $m_{\rm el}$
\lonel, \solut\ to $m\approx
m_{\rm mag}$. Accordingly, the interface tension $\alpha$, \albub\ changes from
$\alpha\simeq {1\over g\beta^3}$ to $\alpha\simeq {1\over g^2\beta^3}$.
This can be easiest seen by computing the contribution of
the potential term to the effective action \sbeff, using the fact
that $q\simeq q(mz)$ and $m\approx m_{\rm mag}$. Higher terms
in the effective action give subleading contributions
to the interface tension.  The free
energy of the domain wall seems to be much larger than previously believed,
and is in accord with the expected non -- perturbative behavior of
asymptotically free gauge theory.

To actually prove the above assertion one would need to derive the
effective action
for finite $q$ and verify that it admits a finite action domain wall solution
with the above described asymptotics.
Due to the non -- perturbative nature of $m_{\rm mag}$ and other problems
this seems difficult at present.

\newsec{Conclusion}

The main purpose of this paper was to study the properties of the effective
action for Wilson loops winding around compact time in different finite
temperature gauge theories. The main results obtained are:

1) We have calculated the quadratic term in the
effective action for the Wilson loop $W_n$ to one loop
order in the gauge coupling constant, and have outlined the calculation of
higher order (in $W$) terms. We found that the inverse propagator
of Wilson loops $G^{(2)}$
\seff, \gtol\ contained a branch cut which was interpreted
as due to pair production of constituents in the external $W_n$ field.
The improved understanding of the dynamics of Wilson loops was then used
to reconsider two recent proposals to apply high temperature perturbation
theory to different physical problems.

2) We have discussed the idea \pol\ (see also \kut)
that one can use perturbative techniques in QCD to
deduce properties of the confining phase, ``analytically
continued'' to high temperature. We argued that the analytic
structure of the quadratic term in the effective action
for Wilson loops that we found does not support such an interpretation
of the perturbative calculations. The inverse propagator in the
confining phase is not expected to exhibit any singularities in momentum
space, to leading order in $1/N$, while such singularities do appear
in the perturbative results.

3) We found that certain domain walls between different
vacua with broken $Z_N$ symmetry that were extensively studied
in recent literature \domw\ -- \cosm, \diff\ --  \smilga\
are modified significantly due to infrared effects.
In particular, we argued that the free energy of such
domain walls behaves like $1/g^2$ as opposed to $1/g$
as previously believed.

There are many natural extensions of this work.
One would like to extend the results obtained here
to higher orders in the gauge coupling. In order to study
the large $N$ deconfinement (Hagedorn) transition
one needs, as we saw, to calculate the Green's
functions in \seff\ to all orders in $g$.
This may be feasible in toy models of lower dimensional Yang Mills
theory \tba, where one may use extensions of our techniques
to calculate the Hagedorn temperature and perhaps to explicitly see
the change in the analytic structure of the action \seff\
between the deconfined and confining phases.
The physics of the branch cuts in the Wilson loop propagator found above,
and in particular their relation to ones that appear in the true
high temperature vacuum (at different $k^2$ in general), also needs
to be understood much better.

It would be interesting to understand what are the implications
of the larger free energy of domain walls found here for the
cosmological scenarios described in the literature \domw, \cosm.
Also, a better understanding of the dynamics of Wilson loops
may suggest ways to study the disintegration of fundamental
strings into their purported constituents suggested in
\ref\joe{M. Natsuume and J. Polchinski,
Santa Barbara preprint NSF-ITP-94-19, (Feb. 1994), hep-th/9402156.}.
In particular, it would be interesting to obtain a similar picture
to that found here for unified strings above the Hagedorn
temperature.

\bigbreak\bigskip\centerline{{\bf Acknowledgements}}\nobreak

D.K. thanks the Aspen Center for Physics for hospitality while this work was
concluded.  This work was partially supported by a DOE OJI grant and
the L. Block foundation.  J.B. is supported by a NSF Graduate Fellowship.

\listrefs

\end